\documentclass[a4paper]{jpconf}
\usepackage{graphicx}
\begin{document}
\title{Occulting Light Concentrators in Liquid Scintillator Neutrino Detectors}

\author{Margherita Buizza Avanzini$^{a}$, Anatael Cabrera$^{b}$, Stefano Dusini$^{c}$, Marco Grassi$^{d}$, Miao He$^{d}$, Wenjie Wu$^{d,e}$}

\address{$^{a}$CNRS/IN2P3/LLR , Palaiseau, France; $^{b}$CNRS/IN2P3/APC and LNCA Laboratories, Paris, France; $^c$INFN-Padova, Padova, Italy; $^d$Institute of High Energy Physics, Beijing, China; $^e$China Wuhan University, Wuhan, China.}

\ead{buizza@llr.in2p3.fr}

\begin{abstract}
The experimental efforts characterizing the era of precision neutrino physics revolve around collecting high-statistics neutrino samples and attaining an excellent energy and position resolution. 
Next generation liquid-based neutrino detectors, such as JUNO, HyperKamiokande, etc, share the use of a large target mass, and the need of pushing light collection to the edge for maximal calorimetric information. Achieving high light collection implies considerable costs, especially when considering detector masses of several kt. 
A traditional strategy to maximize the effective photo-coverage with the minimum number of PMTs relies on Light Concentrators (LC), such as Winston Cones. In this paper, the authors introduce a novel concept called Occulting Light Concentrators (OLC), whereby a traditional LC gets tailored to a conventional PMT, by taking into account its single-photoelectron collection efficiency  profile and thus occulting the worst performing portion of the photocathode. 
Thus, the OLC shape optimization takes into account not only the optical interface of the PMT, but also the maximization of the PMT detection performances. The light collection uniformity across the detector is another advantage of the OLC system. 
By considering the case of JUNO, 
we will show OLC capabilities in terms of light collection and energy resolution.

\end{abstract}

\section{Introduction}
The traditional goal of non-imaging light concentrators (LCs) has always been to maximize the collected light (see for instance \cite{bib:LC-sno} and \cite{bib:LC-others}).
This is particularly true for detectors with a limited ($\sim$ 30\%) geometrical coverage, such as SNO, CTF, Borexino, etc.
Moreover, in a liquid based detector the amount of collected light depends on the vertex position, due to light absorption and scattering.
By limiting the field of view of the PMT, LCs reduce the amount of collected light for events at large radii, thus helping to increase the energy reconstruction uniformity. 
In detectors as JUNO and RENO-50, planning very high geometrical coverages (around 70\%),  
LC remains a valid tool
also against another non-homogeneity effect. Actually, in large PMTs the photoelectron (p.e.) Detection Efficiency (DE) is uneven throughout their surface - the PMT edge being the worst-performing region. 
The dispersive effects induced by such a non-homogeneous DE affect the total number of collected p.e., worsening significantly the detector-level energy resolution when compared to the intrinsic stochastic component. Standard LCs can become Occulting Light Concentrator (OLC) with the aim of reducing PMT-related dispersive effects while maximizing light collection and making the detector response more uniform.

\section{OLCs in a gigantic (15\,kt) liquid scintillator detector}
\begin{figure}
\centering
\includegraphics[width=0.28\textwidth]{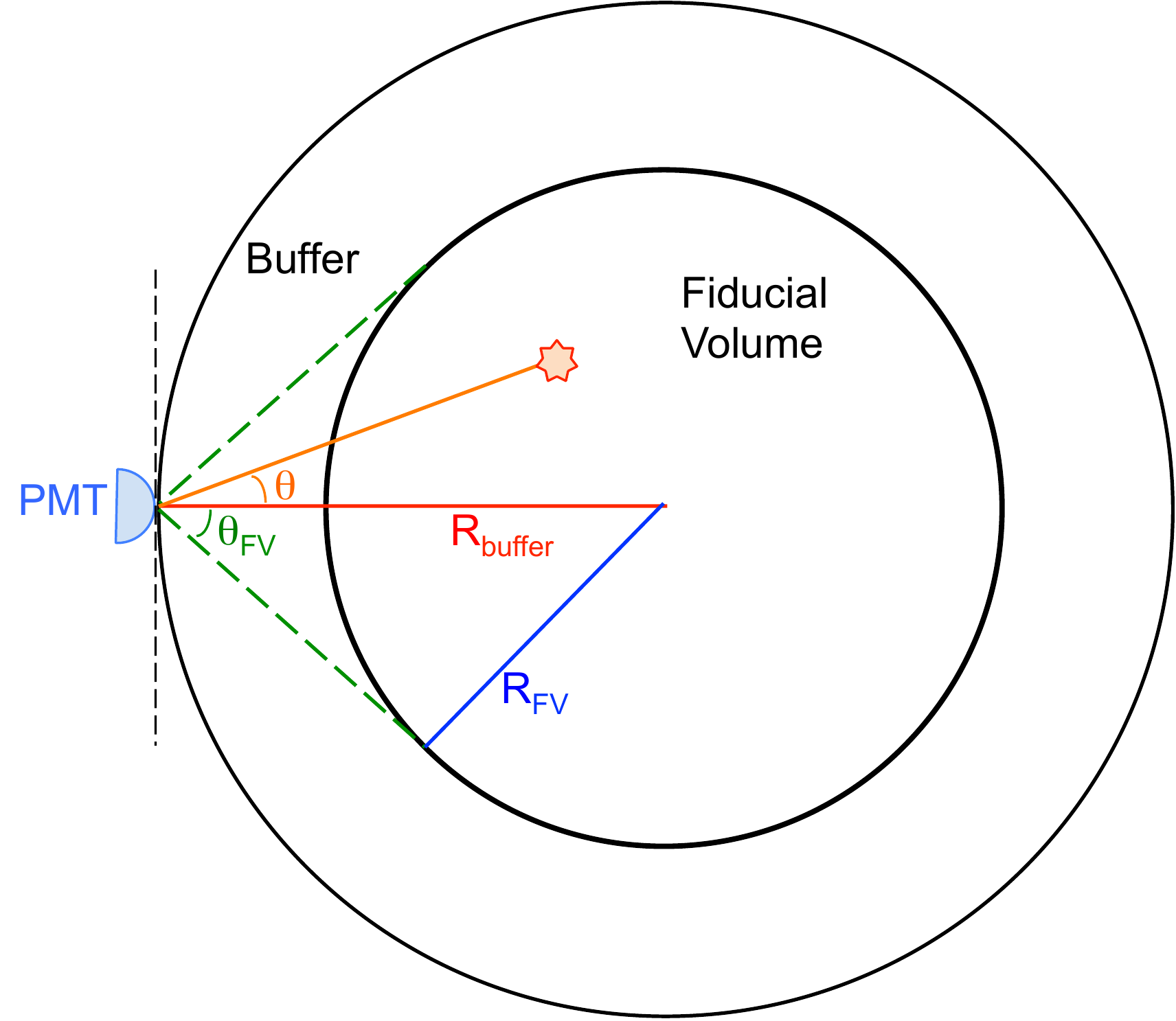}
\includegraphics[width=0.33\textwidth]{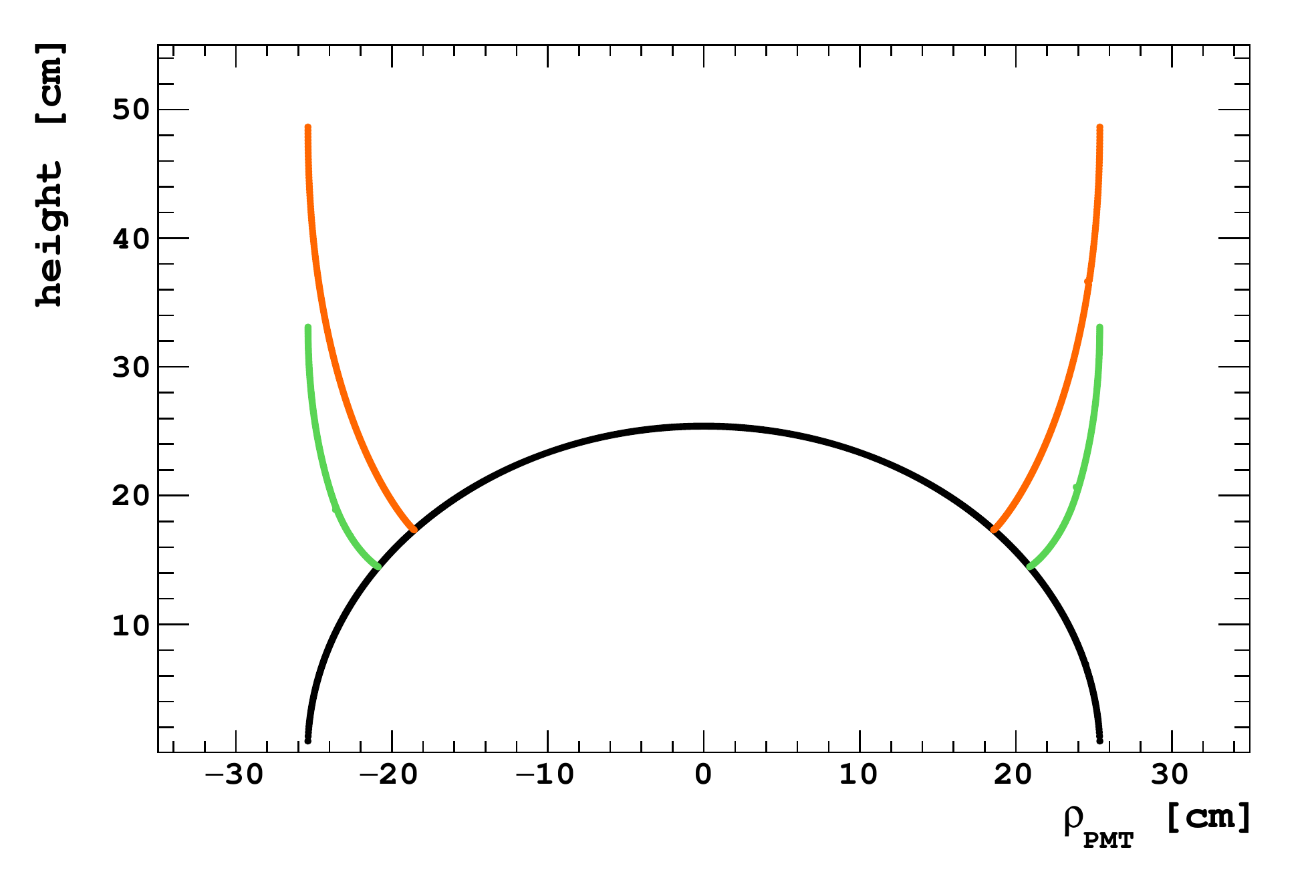}
\includegraphics[width=0.35\textwidth]{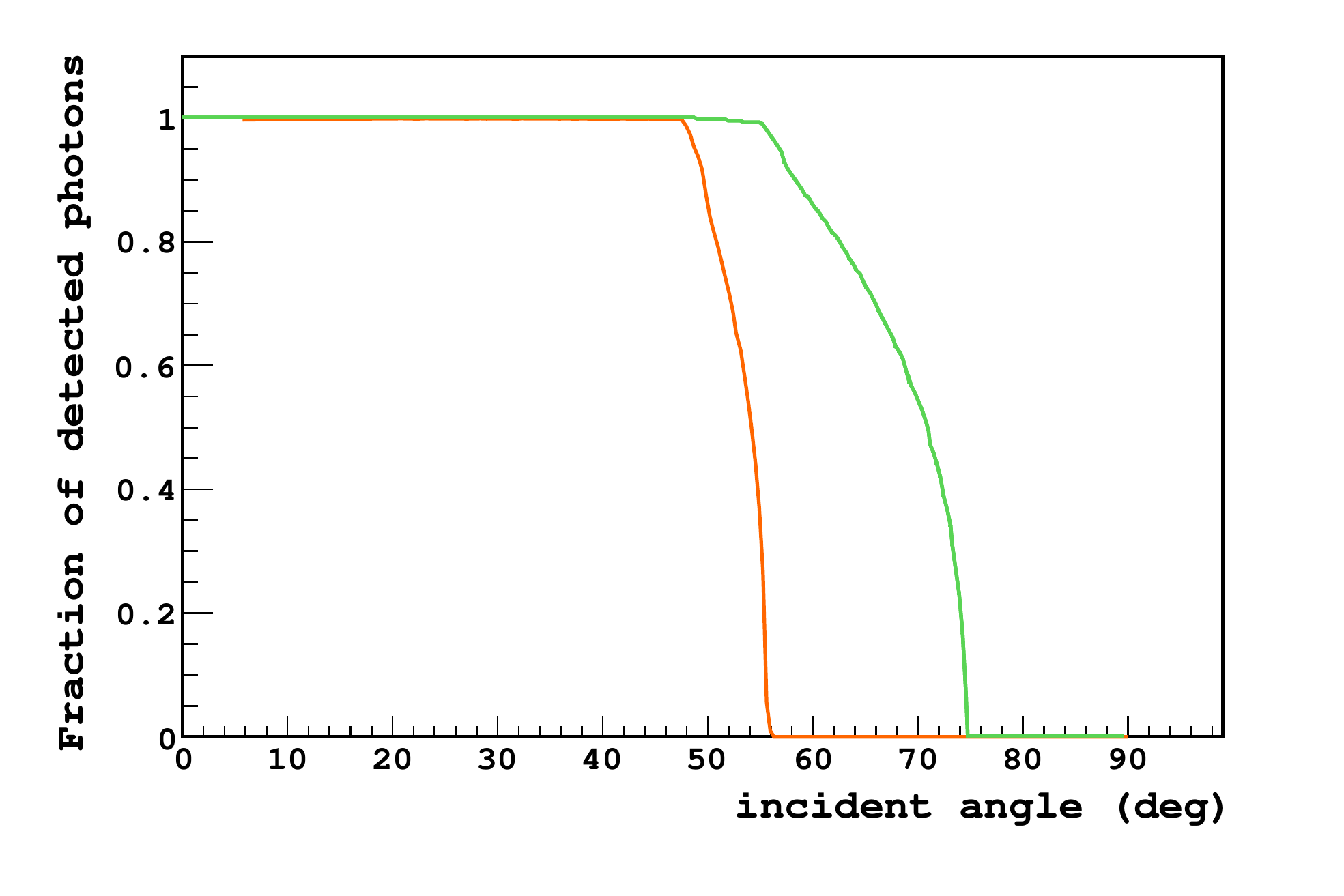}
\caption{\textit{Left: }Geometry of the detector Fiducial Volume and of the PMT. $\theta_{FV}$ corresponds to the maximal photon incident angle at the PMT level. \textit{Center:} 2D profile of OLC$_{1}$ (orange) and OLC$_2$ (green), as obtained by requiring the maximum radius of the LC not to exceed the PMT radius via the Tangent Ray Method. 
\textit{Right: }Acceptance of the optical module obtained with a 3D OLC$_{1}$ (orange) and OLC$_2$ (green). In our case, $\theta_{FV}$ is about 55$^{\circ}$.}
\label{fig:FV}
\end{figure}

An ideal two-dimensional Compound Tangential Concentrator (CTC) is designed in order to transmit to the photocathode all the light incident at the entrance aperture with an angle $<\theta_i$ and to avoid the transmission of all the light with incident angle $>\theta_i$ \cite{bib:LC-sno}. Thus, $\theta_i$ is a CTC construction parameter and defines the maximum accepted incident photon angle. In 3D, where the CTC is obtained by rotating the 2D profile around the symmetry axis, the perfect performances in light transmission are slightly degraded for angles $<\theta_i$. We call $\theta^*$ the incident angle where this degradation starts
and $\theta_{FV}$ the maximal photon incident angle at the PMT level (see Figure \ref{fig:FV}). We consider two extreme configurations: OLC$_1$, obtained by requiring $\theta_i = \theta_{FV}$ and OLC$_2$, by requiring $\theta^* = \theta_{FV}$. We simulate a detector Fiducial Volume (FV) as a finite spherical light source of 16\,m radius and the PMT as a semi-sphere of 25.4\,cm of radius, 4\,m away from the edge of the FV. 
\begin{figure}
\centering
\includegraphics[width=0.28\textwidth]{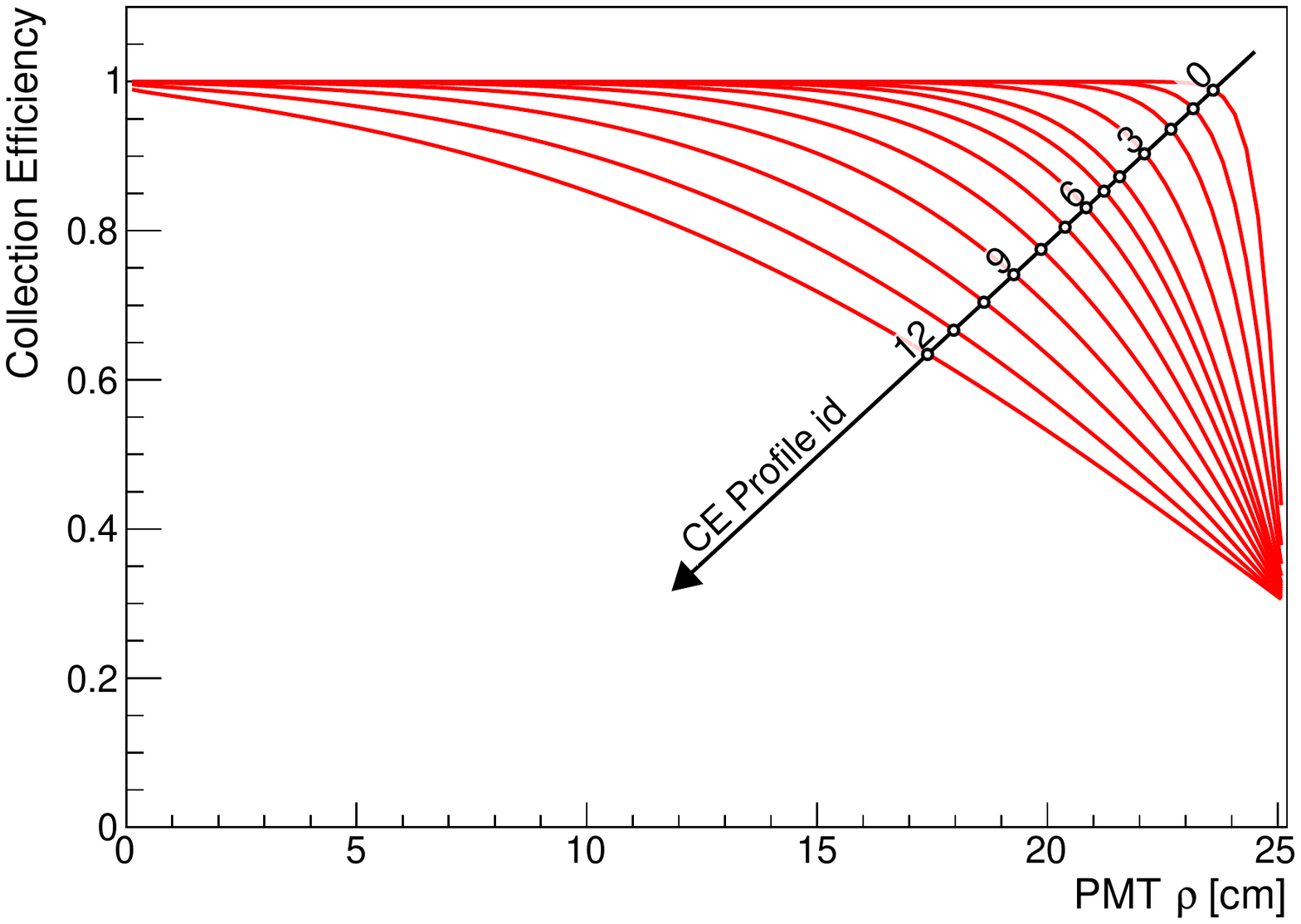}
\includegraphics[width=0.35\textwidth]{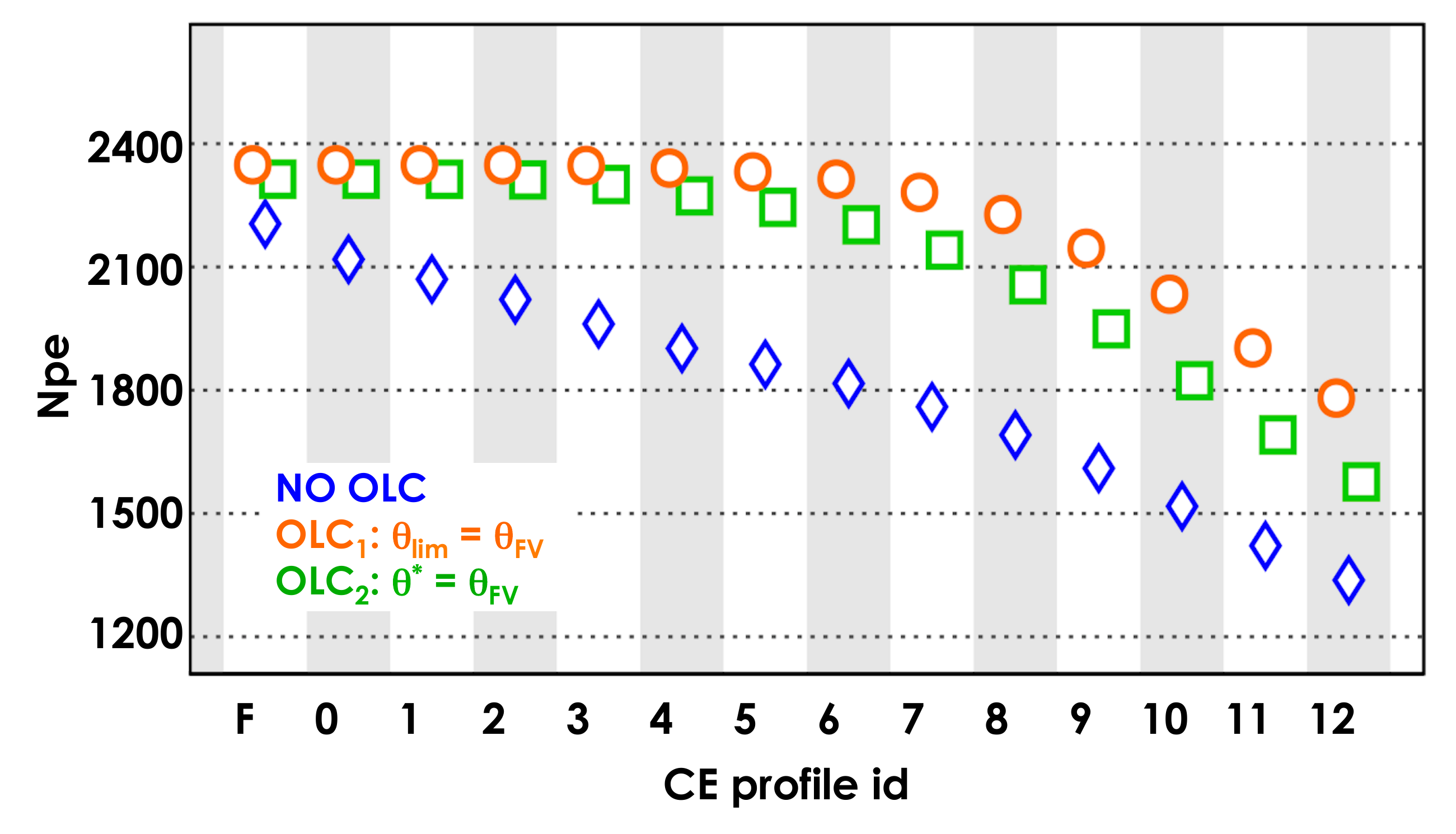}
\includegraphics[width=0.35\textwidth]{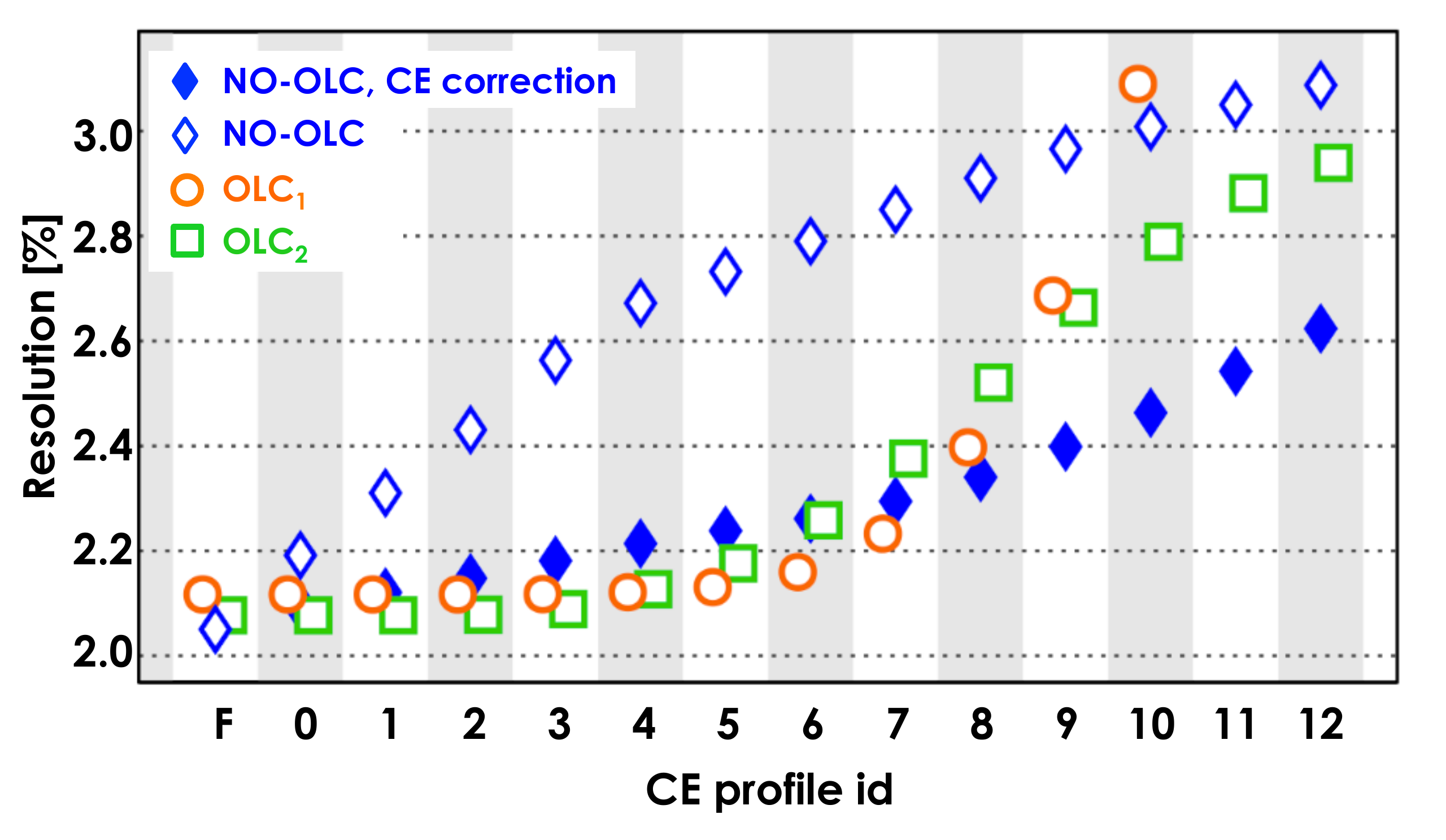}
\caption{\textit{Left: } Collection Efficiency Profiles as a function of $\rho_{PMT}$. \textit{Center:} Light Yield for different CE profiles. \textit{Right: }Detector Resolution for different CE profiles.}
\label{fig:ce_studies}
\end{figure}
Last panel of Figure \ref{fig:FV} shows the acceptances of the two 3D OLC designs: 
only OLC$_2$ has been configured in order to maintain the maximal acceptance up to $\theta_{FV}$. 
We consider 12 possible CE profiles as a function of the distance from the PMT axis ($\rho_{PMT}$), as shown in Figure \ref{fig:ce_studies} (left). 
A non-flat CE profile acts on the detector resolution by reducing the number of detected photons (increasing the stochastic resolution), as clearly shown by blue dots in central panel of Figure \ref{fig:ce_studies}. Moreover, it introduces a dependence on the event vertex position, resulting in an additional non-stochastic resolution term, as show in the right panel of Figure \ref{fig:ce_studies} (cf. empty and filled dots). OLC focuses the light meant to hit the external crown of the photocathode (large $\rho$, close to the PMT equator) towards the central region of the photocathode, effectively mitigating the dispersive behaviour of the CE profiles. This is shown again in the central and right panel of Figure \ref{fig:ce_studies} for OLC$_1$ (orange) and OLC$_2$ (green).
Plots in Figure \ref{fig:ce_studies} are obtained by applying a radial non-uniformity correction; we assume either perfect knowledge of CE profiles (filled dots) or flat CE (empty dots). For CE profiles up to number 7, OLCs clearly improve the energy resolution. Results are comparable with those obtained by full knowledge of CE profile. 
Thanks to its tallest shape, OLC$_1$ not only collects more light 
in the case of a flat CE, but it also manages to have the best light collection across all the CE models.
However, in terms of energy resolution, OLC$_2$ has better performance than OLC$_1$ also for extreme CE profiles, due to its peculiar acceptance curve.

\section{OLCs in JUNO}
\begin{figure}
\centering
\includegraphics[width=0.99\textwidth]{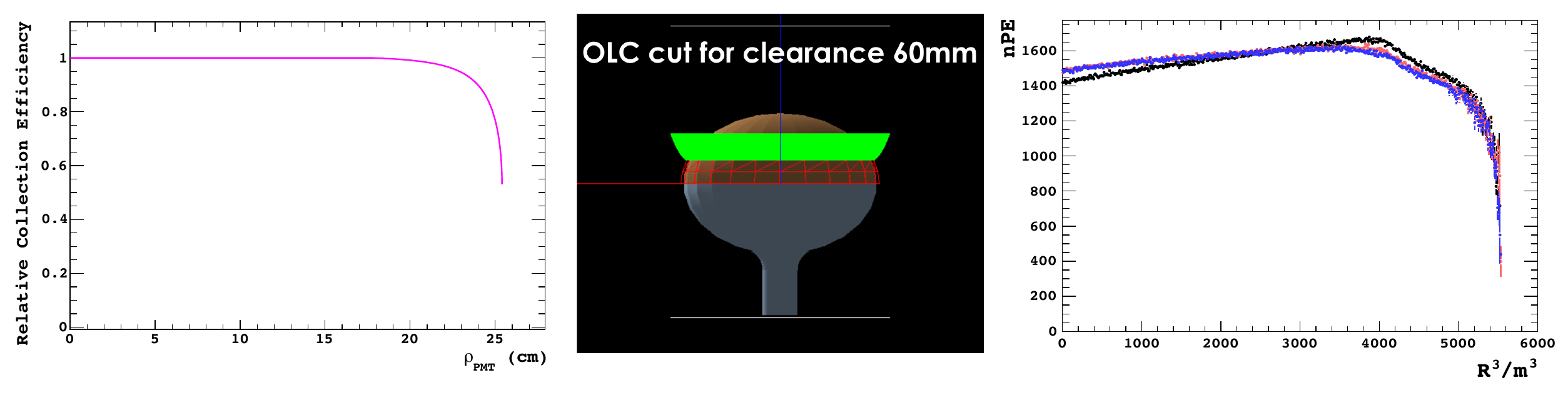}
\caption{\textit{Left: }CE profile as a function $\rho_{PMT}$ as measured for JUNO PMTs. \textit{Center:} JUNO OLC profiles (cut CTC for clearance between two PMTs of 60\,mm). \textit{Right: } Number of collected p.e. as a function of the detector radius in the case without OLC (black) and with OLC designed for a PMT clearance of 45\,mm (red) and 60\,mm (blue). Assumed OLC reflectivity is 0.9.}
\label{fig:juno}
\end{figure}
The largest (20\,kt) liquid scintillator detector currently under construction, JUNO \cite{bib:juno}, will be equipped with about 18\,k 20'' PMTs, with the aim to reach about 75\% of geometrical coverage. The main goal of JUNO is to determine the neutrino Mass Hierarchy by measuring the energy spectrum of $\overline\nu_e$ coming from nuclear reactors at 52\ km far from the detector. To achieve this goal, an unprecedented 3\% energy resolution at 1\,MeV is required. The JUNO collaboration is presently considering to implement OLCs in the detector design, in order to: (1) occult the PMT edge, where the CE decreases; (2) improve the uniformity in the light collection across the detector; (3) recover good light in case the total number of PMTs has to be reduced.
The CE profile has been measured on a sub-sample of JUNO PMTs. The analytical behavior as a function of $\rho_{PMT}$ is reported in Figure \ref{fig:juno} (left), showing an almost constant CE up to $\rho_{PMT}\sim$ 24\,cm. The OLC profile has thus been tailored on JUNO PMT, via the CTC method, in order to occult the photocathode area at $\rho_{PMT}>$ 24\,cm, for R$_{FV}=$ 17.2\,m, R$_{buffer}=$ 19.5\,m and by asking $\theta_i = \theta_{FV}$. Since the complete CTC shape does not fit JUNO geometrical constraints, it must be cut in order to avoid overlapping with near-by OLCs, as shown in Figure \ref{fig:juno} (middle).  The reference clearance between two PMTs is 25\,mm. Due to mechanical constraints, larger clearances are begin considered. To increase the clearance also means to reduce the total number of PMTs. As shown in Figure \ref{fig:juno} (right), OLCs help to recover light when increasing the clearance. Moreover, the light collection becomes more uniform across the detector volume.  
Thanks to their occulting action and to the uniformity in the light detection, also the energy resolution improves. 
In particular, when requiring a larger clearance, OLC allow to recover the energy resolution level of the standard configuration (no-OLC and 25\,mm clearance), thus avoiding any degradation in the energy reconstruction.

\end{document}